\newcommand{\be}{\begin{equation}}
\newcommand{\bea}{\begin{eqnarray}}
\newcommand{\ee}{\end{equation}}
\newcommand{\eea}{\end{eqnarray}}
\newcommand{\nn}{\nonumber}

\newcommand{\qa}{\alpha}
\newcommand{\qb}{\beta}
\newcommand{\qg}{\gamma}

\newcommand{\qd}{\delta}
\newcommand{\qD}{\Delta}
\newcommand{\qe}{\varepsilon}

\newcommand{\qh}{\eta}

\newcommand{\qr}{\rho}
\newcommand{\qs}{\sigma}

\newcommand{\qps}{\psi}

\newcommand{\qj}{\psi}

\newcommand{\qO}{\Omega}

\newcommand{\tr}{{\rm tr}\,}
\newcommand{\Tr}{{\rm Tr}\,}

\newcommand{\dagg}{^{\dag}}

\renewcommand{\Re}{{\rm Re}\,}
\renewcommand{\Im}{{\rm Im}\,}
\newcommand{\rd}{{\rm d}}

\newcommand{\CC}{{\mathbb C}}

\newcommand{\EE}{{\mathbb E}}

\newcommand{\RR}{{\mathbb R}}

\newcommand{\one}{\mathbb{1}}

\newcommand{\cC}{{\mathcal C}}
\newcommand{\cD}{{\mathcal D}}

\newcommand{\cF}{{\mathcal F}}

\newcommand{\cK}{{\mathcal K}}

\newcommand{\cS}{{\mathcal S}}

\newcommand{\isdef}{\stackrel{\rm def}{=}}

\newcommand{\ket}[1]{| #1 \rangle}
\newcommand{\bra}[1]{\langle #1 |}
\newcommand{\ketbra}[1]{\ket{#1}\bra{#1}}

\newcommand{\acc}{{\rm acc}}
\newcommand{\rej}{{\rm rej}}
\newcommand{\EPR}{{\rm EPR}}

\documentclass[sn-mathphys,Numbered]{sn-jnl}

\usepackage{graphicx}%
\usepackage{multirow}%
\usepackage{amsmath,amssymb,amsfonts}%
\usepackage{amsthm}%
\usepackage{mathrsfs}%
\usepackage[title]{appendix}%
\usepackage{xcolor}%
\usepackage{textcomp}%
\usepackage{manyfoot}%
\usepackage{booktabs}%
\usepackage{algorithm}%
\usepackage{algorithmicx}%
\usepackage{algpseudocode}%
\usepackage{listings}%
\usepackage{bbold}%
\usepackage{makecell}%
\usepackage{enumitem}%
\usepackage{array}
\usepackage{tabularray}

\newtheorem{theorem}{Theorem}[section]

\newtheorem{lemma}[theorem]{Lemma}
\newtheorem{definition}[theorem]{Definition}

\begin{document}

\title[Article Title]{Authentication of Continuous-Variable Quantum Messages}


\author*[1]{\fnm{Mehmet H\"{u}seyin} \sur{Temel}}\email{m.h.temel@tue.nl}

\author[2]{\fnm{Boris} \sur{\v{S}kori\'{c}}}\email{b.skoric@tue.nl}

\affil[1]{\orgdiv{Electrical Engineering Dept.}, \orgname{Eindhoven University of Technology}, \orgaddress{\city{Eindhoven}, \postcode{5600MB}, \country{The Netherlands}}}

\affil[2]{\orgdiv{Mathematics and Computer Science Dept.}, \orgname{Eindhoven University of Technology}, \orgaddress{\city{Eindhoven}, \postcode{5600MB}, \country{The Netherlands}}}

\abstract{
We introduce the first quantum authentication scheme for continuous-variable states.
Our scheme is based on trap states, and is an adaptation of a discrete-variable scheme 
by Broadbent et al. \cite{broadbent2013quantum}, but with more freedom in choosing the number of traps.

We provide a security proof, mostly following the approach of Broadbent and Waine\-wright~\cite{broadbent2016efficient}.
As a necessary ingredient for the proof
we derive the continuous-variable analogue of the Pauli Twirl.
}




\maketitle

\section{Introduction}
With the rapid advancement of quantum technologies and the increasing deployment of quantum communication systems, new protocols for the secure transmission of quantum information have been proposed. While Quantum Key Distribution (QKD)~\cite{bennett1984quantum, alleaume2010quantum, diamanti2016practical, cao2022evolution} is the most widely known application—using quantum systems to establish shared classical keys for classical encryption—quantum cryptography provides a broader set of protocols designed to protect quantum data itself. These include quantum encryption~\cite{ambainis2000private, AMTW2000, BR2003, bradler2005continuous, jeong2015gaussian}, quantum secret sharing \cite{liu2023experimental, cleve1999share}, and quantum message authentication~\cite{barnum2002authentication, broadbent2016efficient, aharonov2017interactive}.
\subsection{Quantum Authentication}
Message authentication is a fundamental task in cryptography that enables a receiver to verify whether a message has been tampered with during transmission, and whether it originates from the claimed sender. 
By enabling tamper detection and origin verification, 
quantum authentication serves as a critical building block for advanced cryptographic protocols such as quantum one-time programs \cite{broadbent2013quantum} and secure multiparty quantum computation \cite{ben2006secure, DNS12, dulek2020secure}.

A quantum authentication scheme works with a classical symmetric key and 
consists of two keyed procedures: encoding (or encryption) and decoding (or decryption). 
The sender encodes the quantum message using the key.
The recipient gets a quantum state and decodes it using the same key;
attempts to forge or manipulate the quantum message are detected with high probability.

The first Quantum Authentication Scheme (QAS) was introduced by Barnum et al. \cite{barnum2002authentication}, 
where they also provided security definitions.
One of the key results of their work was that any QAS must encrypt the quantum message. Their construction relied on purity-testing codes derived from quantum error-correcting codes (QECCs). It was later shown that such purity-testing codes can also satisfy universal composability \cite{hayden2016universal}.

The original security definition has since been strengthened by more robust proposals \cite{oppenheim2005, hayden2016universal, garg2017new}. 
It was shown that partial or even complete key reuse is possible, depending on the amount of key leakage \cite{oppenheim2005, portmann2017quantum, dulek2018trapcode}. 
A variety of QAS constructions have been proposed, based on polynomial codes \cite{ben2006secure, aharonov2017interactive}, Clifford codes \cite{aharonov2017interactive}, threshold codes \cite{dulek25}, and trap codes \cite{broadbent2013quantum, broadbent2016efficient}.

\subsection{Trap Code-Based Quantum Authentication}

In this paper, we focus on trap code-based quantum authentication, first introduced in \cite{broadbent2013quantum} and later refined with a more efficient security proof in \cite{broadbent2016efficient}. 
The main idea behind trap codes is to insert dummy states—referred to as traps—into the quantum message (which has already been encoded using a QECC), and then apply a secret permutation and One-Time Pad encryption. 
The traps are used to detect any tampering by an adversary.



\subsection{Contribution}
All of the quantum authentication schemes referred above are designed for discrete-variable (DV) quantum states. 
In contrast, continuous-variable (CV) quantum message authentication remains relatively unexplored. 
CV systems are particularly attractive for practical implementations due to their compatibility with existing optical communication infrastructure.

In this work, we introduce the first quantum authentication scheme for CV quantum states. 
Our construction is an adaptation of the DV trap code-based QAS proposed by Broadbent, Gutoski, and Stebila \cite{broadbent2013quantum}, and it follows the proof technique of Broadbent and Wainewright \cite{broadbent2016efficient}. 
While our construction and security proof take the same global steps as \cite{broadbent2016efficient}, 
the differences between DV and CV 
lead to several nontrivial features, e.g.
the necessity to allow a small probability of error in the verification step;
fine tuning scheme parameters in order to obtain properly matched step-like functions;
a CV analogue of the Pauli twirl.

Our contributions are:
\begin{itemize}
\item 
We propose the first quantum authentication scheme for CV states.
In contrast to existing DV schemes,
our construction allows for a variable number of trap states.
\item 
We provide a security proof for the proposed scheme, adapting and extending techniques from the DV setting to the CV setting.
\item 
We introduce the notion of a \emph{CV Twirl}, an analogue to the Pauli Twirl.
\end{itemize}

\section{Preliminaries}
\label{sec:prelim}

\subsection{Notation}
\label{sec:notation}
We use standard notation from quantum information theory. 
Quantum states are represented by density operators (positive semi-definite, trace-one operators) acting on Hilbert spaces, and we write them as \(\rho, \sigma\), etc., with subscripts indicating associated registers.
The identity operator is denoted by \(\one\), and the partial trace over a subsystem \(A\) is \(\Tr_A\). 
We write \(\|A\|_1\) for the 1-norm, and 
$\| \qr-\qs \|_{\rm tr}=$
\(  \frac12\|\rho - \sigma\|_1\) is the trace distance between two quantum states. 
The quadrature operators \(\hat x\) and \(\hat p\) correspond to the position and momentum observables, respectively.
A coherent state \(\ket{z}\) is defined as the eigenstate of the annihilation operator 
$\hat a=\frac{\hat x+i\hat p}{\sqrt2}$
with complex eigenvalue \(z \in \mathbb{C}\). 

We will use the notation $\ket X$ for a single-mode squeezed state that is narrow in the direction of the x-quadrature, and centered on zero.
The Wigner function of such a squeezed state is Gaussian, with variance $e^{-r}$ in the $x$-direction and
$e^r$ in the $p$-direction; the $r$ is called the squeezing parameter.
Similarly we define the single-mode squeezed state $\ket P$.

The displacement operator \(D(\qb)=e^{\qb\hat a\dagg-\qb^*\hat a}  \) shifts the phase space of a mode. 
The action on a coherent state is given by $D(\qb)\ket z = \ket{z+\qb}$.
Quantum One-Time Pad (QOTP) encryption for CV is achieved by applying a secret displacement 
chosen from a wide complex Gaussian distribution, for each mode independently.


A CV Quantum Error Correcting Code (CV-QECC) is called an \([[n, 1, d]]\) code if it 
encodes one mode to $n$ modes and 
is capable of correcting arbitrarily large displacements in up to \(t = \lfloor (d-1)/2 \rfloor\) out of $n$ modes.



\subsection{Security definitions and useful lemmas}

\begin{lemma}(See e.g. \cite{walls1994quantum})
The displacement operation  $D$ satisfies the property
\be
\label{displacement}
    D(\beta) D(\gamma) = e^{i\, \Im(\beta \bar{\gamma})} D(\beta + \gamma).
\ee
\end{lemma}

We use the definition of a quantum authentication scheme given by Broadbent et al.\cite{broadbent2016efficient},
but with a small modification: we allow for a small probability that a decoding error occurs.\\

\begin{definition}[Quantum message authentication scheme]
\label{def:QAS}
A quantum authentication scheme (QAS) is a polynomial-time set of encryption and decryption channels $\{\mathcal{E}_k^{M \rightarrow C}, \mathcal{D}_k^{C \rightarrow MF}) \mid k \in \mathcal{K} \}$ where $\mathcal{K}$ is the set of possible keys, $M$ is the input system, $C$ is the encrypted system, and $F$ is a flag system indicating either acceptance $\ket{acc}$ or rejection $\ket{rej}$ such that
\be
    \forall_{\qr_M}\quad
    \Big\| (\mathcal{D}_k \circ \mathcal{E}_k)(\rho_M) - \rho_M \otimes \ket{\acc}\bra{\acc}_F\Big\|_{\rm tr} \leq \qe_{\rm dec},
\ee
where $\qe_{\rm dec}$ is a small decoding error probability.
\end{definition}
We allow the message register $M$ to be entangled with a reference system $R$ that belongs to the adversary.
The input to the scheme is expressed as a joint quantum state $\rho_{MR}$.

The adversary applies a joint unitary $U_{CR}$ on the encoded message and the reference system. 
For a fixed key $k$, the corresponding real-world quantum channel is defined as
\be
\mathcal{C}_k^{MR \to MRF} :\quad
\rho_{MR} \mapsto (\mathcal{D}_k \otimes \one_R) \left( U_{CR}\; (\mathcal{E}_k \otimes \one_R)(\rho_{MR})\; U_{CR}^\dagger \right).
\ee

The security definition relies on comparing this real-world channel with an idealized simulator which has access only to the ideal functionality. The ideal functionality either accepts the message by outputting message register M, or rejects it by outputting a fixed dummy state $\Omega_M$. The simulator may also modify the reference system $R$. The idealized process can be expressed as ideal channel $\mathcal{F}$, 
\bea
    \mathcal{F}^{MR \to MRF}: \quad & \rho_{MR} \mapsto & (\one_M \otimes \mathcal{U}_R^{\acc}) (\rho_{MR}) \otimes \ket{\acc}\bra{\acc} 
    \nn\\ &&
    + \Omega_M \otimes \tr_M\big[(\one_M \otimes \mathcal{U}_R^{\rej})(\rho_{MR}) \big]  \otimes \ket{\rej}\bra{\rej},
\label{Fgeneral}
\eea
 where for each attack $U_{CR}$ there exists two CP maps $\mathcal{U}_R^{\acc}$ and $\mathcal{U}_R^{\rej}$ acting only on the reference system $R$, satisfying $\mathcal{U}_R^{\acc} + \mathcal{U}_R^{\rej} = \one_R$.\\

\begin{definition}[Security of quantum message authentication \cite{broadbent2016efficient} ] 
\label{def:security}
Let $\{(\mathcal{E}_k^{M \to C}, \mathcal{D}_k^{C \to MF}) \mid k \in \mathcal{K}\}$ be a quantum message authentication scheme.
The scheme is $\qh$-secure if for all attacks there exists a simulator $\mathcal{F}$ such that
\be
\forall \rho_{MR} \quad\quad
\Big\| \frac{1}{|\cK|} \sum_{k\in\cK}\mathcal{C}_k(\rho_{MR}) - \mathcal{F}(\rho_{MR})  \Big\|_{\rm tr} \leq \qh,
\ee
where the simulator has access only to the ideal functionality of the scheme. 
\end{definition}



\section{Trap Code CV Quantum Authentication Scheme}
\label{sec:construction}
We construct our CV quantum authentication scheme by adapting the trap code-based DV construction of Broadbent et al. \cite{broadbent2013quantum} to CV quantum states. 
The encryption process begins with encoding the message modes using a quantum error-correcting code (QECC). Subsequently, two sets of trap modes are inserted. 
The entire set of modes is permuted and then encrypted using a CV quantum one-time pad. 
The decoding process reverses the encoding steps: 
the received state is first decrypted and de-permuted, after which the integrity of the trap modes is verified. 
If the trap modes are intact, the message modes are decoded using the QECC, and `accept' is flagged. 
If the trap modes are not intact, a `reject' is flagged, the message state is discarded (traced out), and a dummy message is output instead.\\ 

\noindent                                                                 
\underline{\textbf{Encoding}}\\
The encoding process, denoted by $\mathcal{E}^{M \to C}_{k}$, takes as input the single-mode message state $\rho_M$. 
A CV QECC with parameters $[[n, 1, d]]$ is applied to $\rho_M$, encoding it 
to ${\rm Enc}(\qr_M)$ which consists of $n$ modes. 
The QECC is able to correct displacements in $\leq t$ modes, where $d = 2t +1$.

After encoding, $z$ states squeezed in the $x$-quadrature and $z$ states squeezed in the $p$-quadrature are appended to the encoded message, forming a system of $n + 2z$ modes. 
For proof-technical reasons we set $2z>n$.
These squeezed states are denoted as $\ket{X}$ and $\ket{P}$, respectively, and act as traps.
They are centered on zero and have squeezing parameter~$r$.
The entire set of modes is then permuted according to a secret key $k_1$.
Finally, a QOTP is applied according to a secret key $k_2\in\CC^{n+2z}$ drawn
from a Gaussian distribution with variance  $\qD^2 \gg 1$.
We write $k=(k_1,k_2)$.
The output Hilbert space has $n+2z$ modes.

The QAS encoding is expressed as: 
\begin{equation}
    \mathcal{E}^{M \to C}_{k} : \rho_M \mapsto \rho_C\quad\quad
    \rho_C=D_{k_2} \pi_{k_1} \Big( \text{Enc}(\rho_M) \otimes \ket{X}\bra{X}^{\otimes z} \otimes \ket{P}\bra{P}^{\otimes z} \Big) \pi_{k_1}^\dagger D_{k_2}^{\dagger},
\end{equation}
where $\pi_{k_1}$ is the permutation and $D_{k_2}$ is the QOTP displacement operator.\\

\noindent
\underline{\textbf{Decoding}}\\
The decoding $\mathcal{D}^{C \to MF}_{k}$ begins by applying the inverse displacement $D_{k_2}^{\dagger}$ and inverse permutation $\pi_{k_1}^{\dagger}$ to the received cipherstate.
The last $2z$ modes, corresponding to appended squeezed trap states, are then measured using homodyne detection. 
The measurement outcomes are denoted as $(x_i)_{i=1}^z$ and $(p_i)_{i=1}^z$.
We define the following condition for acceptance: 
\begin{equation}
    \forall_{i \in \{1, \dots, z\}} \quad |x_i| \leq \epsilon
    \;\wedge\; |p_i| \leq \epsilon.
\end{equation}
In order to prevent the decoding error probability $\qe_{\rm dec}$ from becoming large,
the values of $r$ and $\epsilon$ are chosen such that $\epsilon\gg e^{-r/2}$.

In case of Accept, QECC-decoding Dec$: \qr_C\to\qr_M$ is applied to the first $n$ modes, and a flag $\ketbra{\text{acc}}$ is appended.
If any trap state fails the condition, the message system $M$ is traced out, and a fixed dummy state $\Omega_M$ is output instead. In this case, the flag $\ketbra{\text{rej}}$ is appended. \\
We define a POVM  $V$ that acts on the trap space and has outcomes $\{\text{acc}, \text{rej}\}$.
The POVM elements are given by
\begin{equation}
    V^{\text{acc}}_\epsilon = 
    \one^{\otimes n} \otimes 
    \Bigg[ \int_{-\epsilon}^\epsilon\!\! \rd x \, \ketbra x \Bigg]^{\otimes z} 
    \otimes 
    \Bigg[ \int_{-\epsilon}^\epsilon\!\! \rd p \, \ketbra p \Bigg]^{\otimes z},
    \quad\quad 
    V^\rej_\epsilon=\one-V^\acc_\epsilon,
\label{POVM_V}
\end{equation}
where $\ket{x}$ is an $x$-quadrature eigenstate and $\ket{p}$ is a $p$-quadrature eigenstate.
The decoding process is expressed as follows, 
\bea
    \mathcal{D}^{C \to MF}_{k}: \rho_C \mapsto 
    \text{Dec}\,  \left(\Tr_{\rm trap}  \sqrt{V^{\acc}_\epsilon} \pi_{k_1}^\dagger D_{k_2}\dagg \rho_C D_{k_2} \pi_{k_1} 
    \sqrt{V^{\acc\dagger}_\epsilon} \right) 
    \otimes\ket{\text{acc}}\bra{\text{acc}} 
    \nonumber\\
    +  \Omega_M \Tr_{M,\rm trap} \left( \sqrt{V^{\rej}_\epsilon} \pi_{k_1}^\dagger D_{k_2}\dagg \rho_C D_{k_2} \pi_{k_1} \sqrt{V^{\rej\dagger}_\epsilon} \right) 
    \otimes \ket{\text{rej} } \bra{\text{rej}}.
\eea

We briefly show that our scheme satisfies Def.\,\ref{def:QAS}.
If there is no noise, each trap state has the following probability of passing the verification:
$\int_{-\epsilon}^\epsilon \! \rd x\; (2\pi e^{-r})^{-1/2} \exp(-\frac{x^2}{2e^{-r}})$
$={\rm Erf}\frac\epsilon{e^{-r/2}\sqrt2}$. Then
\be
    1-\qe_{\rm dec} = \Big({\rm Erf}\frac\epsilon{e^{-r/2}\sqrt2}\Big)^{2z}.
\ee
Given our parameter tuning $e^{-r/2}\ll \epsilon$, the above expression is close to~1.

\section{Security of our Scheme}
\subsection{Approach}

We prove the security of our scheme using a simulator-based approach similar to the one employed by Broadbent and Wainewright \cite{broadbent2016efficient}. 
The main idea is to model a simulator that mimics the ideal functionality of the scheme and then compare it with the real-world execution. 
This comparison establishes the security of our construction by showing that any adversary interacting with the real protocol cannot distinguish it from the ideal case except with negligible probability.

The security proof of \cite{broadbent2016efficient} relies on a simulator that replaces the entire ciphertext $C$ with EPR pairs and permutes them. 
The simulator retains one half of each EPR pair and then it runs the adversary on these EPR states and the reference system~$R$. 
After the attack, the simulator unpermutes the states and performs Bell measurements on the EPR pairs to determine whether they have been tampered with. 
If the EPR pairs remain intact, the simulator outputs “accept”; otherwise, it outputs “reject,” as prescribed by the ideal functionality. 

We adapt the proof method for our CV construction
\begin{itemize}
\item
Qubits are replaced by modes.
\item
Each two-qubit EPR pair is replaced by the two-mode squeezed vacuum.
\item
The DV ``accept'' and ``reject'' projectors become the POVM (\ref{POVM_V}).
\item
The Pauli Twirl is replaced by a CV Twirl. 
\end{itemize}



\subsection{CV Twirl}
\begin{lemma}[CV Displacement Twirl]
\label{CV-twirl} 
Let $D(\cdot)$ be the single-mode displacement operator. For any $\qr$ it holds that
\be
\label{eq:cvtwirl}
\int_{\mathbb{C}}d^2\gamma \frac{1}{2\pi \Delta^2} e^{-\frac{|\gamma|^2}{2\Delta^2}} 
\, D^\dagger(\gamma) D(\beta) D(\gamma) \, \rho \, D^\dagger(\gamma) D^\dagger(\beta') D(\gamma)
=
e^{-2\qD^2 |\qb-\qb'|^2}
D(\beta) \rho D^\dagger(\beta').
\ee
\end{lemma}

Proof: 
From (\ref{displacement}) we have $D(\qb)D(\qg)=e^{\frac{\qb\bar\qg-\bar\qb\qg}2}D(\qb+\qg)$.
Multiplying from the left with $D(-\qg)$ and applying (\ref{displacement}) again yields
$D\dagg(\qg)D(\qb)D(\qg) = e^{\frac{\qb\bar\qg-\bar\qb\qg}2}D(-\qg)D(\qb+\qg)$
$=e^{\frac{\qb\bar\qg-\bar\qb\qg}2} e^{\frac{-\qg(\bar\qb+\bar\qg) +\bar\qg(\qb+\qg)}2} D(\qb)$
$=e^{\qb\bar\qg-\bar\qb \qg} D(\qb)$.
By the same reasoning it holds that 
$D\dagg(\qg) D\dagg(\qb') D(\qg)= e^{-\qb'\bar\qg+\bar\qb' \qg} D\dagg(\qb')$.
We get
\be
    D\dagg(\qg) D(\qb)D(\qg)\;\; \qr\;\; D\dagg(\qg) D\dagg(\qb') D(\qg) = 
    e^{\qg(\bar\qb'-\bar\qb)-\bar\qg(\qb'-\qb)}\; D(\qb) \qr D\dagg(\qb').
\ee
We write $\qg=x+iy$, which gives
$\qg(\bar\qb'-\bar\qb)-\bar\qg(\qb'-\qb) = -2i x  \Im(\qb'-\qb) +2i y \Re(\qb'-\qb)$
and $|\qg|^2=x^2+y^2$. 
The integral over the complex plane becomes two separated Gaussian integrals over $x$ and $y$.
Performing the integrals yields (\ref{eq:cvtwirl}).
\hfill$\square$

{\it Remark.} 
In the DV case the Pauli twirl result is
$\EE_Q Q\dagg P Q \qr Q\dagg {P'}\dagg Q = \qd_{PP'}P\qr P\dagg$, where $P,P',Q$ are $n$-qubit Paulis.
Instead of the Kronecker delta, our result has a Gaussian factor.
Note that the Gaussian factor $e^{-2\qD^2 |\qb-\qb'|^2}$ for $\qD\gg1$
essentially acts as a Dirac delta function which enforces $\qb'=\qb$.



\subsection{Real World Channel}

We introduce shorthand notation
\be
    \qps=\text{Enc}(\rho_{MR}) \otimes \ket{X}\bra{X}^{\otimes z} \otimes \ket{P}\bra{P}^{\otimes z}.
\ee

Using the POVM for the accept case, the real world channel can be expressed as follows: 
\bea
&&\mathcal{C}^{MR \to MRF}: \rho_{MR} \mapsto  \Tr_{\rm trap}\EE_{k_1,k_2} \Bigg\{ \nn \\
    &&\text{Dec}   \Bigg( \sqrt{V^{\acc}_\epsilon} \pi_{k_1}^\dagger D_{k_2}^{\dagger} U_{CR}\Big(D_{k_2} \pi_{k_1} \qps
    \pi_{k_1}^\dagger D_{k_2}^{\dagger}\Big)U^{\dagger}_{CR} D_{k_2} \pi_{k_1} 
    \sqrt{V^{\acc\dagger}_\epsilon} \Bigg) 
    \otimes\ket{\text{acc}}\bra{\text{acc}} 
    \nn \\
    &&+\Omega_M \Tr_{M} \Bigg( \sqrt{V^{\rej}_\epsilon} \pi_{k_1}^\dagger D_{k_2}^{\dagger}  U_{CR}\Big(D_{k_2} \pi_{k_1} \qps
    \pi_{k_1}^\dagger D_{k_2}^{\dagger} \Big) U^{\dagger}_{CR}D_{k_2} \pi_{k_1} \sqrt{V^{\rej\dagger}_\epsilon} \Bigg)   
    \otimes \ket{\text{rej} } \bra{\text{rej}} \Bigg\}.
\eea
The attack is modeled as the unitary $U_{CR}$. 
Analogous to the approach in \cite{broadbent2016efficient},
we expand the attack as $U_{CR} = \int \rd^2\vec\alpha \, \chi(\vec{\alpha}) \, D_C(\vec{\alpha}) \otimes U_{R}^{\vec{\alpha}}$ 
where $\int \rd^2 \vec\alpha$ stands for $\int \rd^2\alpha_1...\rd^2\alpha_{n+2z}$,
and $\int \rd^2\vec\alpha \, |\chi(\vec{\alpha})|^2 = 1$. 
Then the real world channel is given by   
\bea
&&\mathcal{C}^{MR \to MRF}: \rho_{MR} \mapsto  \Tr_{\rm trap}\EE_{k_1} 
    \int \rd^2 \vec\alpha \, \chi(\vec{\alpha}) \, 
    \int \rd^2\vec\alpha' \,\overline{\chi(\vec{\alpha}')}
    \EE_{k_2}\Bigg\{ \nn \\
    &&\text{Dec}   \Bigg( \sqrt{V^{\acc}_\epsilon} \pi_{k_1}^\dagger D_{k_2}^{\dagger} ( D_C(\vec{\alpha}) \otimes U_{R}^{\vec{\alpha}})D_{k_2} \pi_{k_1}\qps\pi_{k_1}^\dagger D_{k_2}^{\dagger} 
     ( D_C(-\vec{\alpha}') \otimes U_{R}^{\vec{\alpha}'\dagger}) D_{k_2} \pi_{k_1} 
    \sqrt{V^{\acc\dagger}_\epsilon} \Bigg) \otimes\ket{\text{acc}}\bra{\text{acc}} \nn \\
    &&+\Omega_M \Tr_{M} \Bigg( \sqrt{V^{\rej}_\epsilon} \pi_{k_1}^\dagger D_{k_2}^{\dagger}  ( D_C(\vec{\alpha}) \otimes U_{R}^{\vec{\alpha}})D_{k_2} \pi_{k_1}\qps\pi_{k_1}^\dagger D_{k_2}^{\dagger}
    ( D_C(-\vec{\alpha}') \otimes U_{R}^{\vec{\alpha}'\dagger})D_{k_2} \pi_{k_1} \sqrt{V^{\rej\dagger}_\epsilon} \Bigg)\otimes \ket{\text{rej} } \bra{\text{rej}} \nn \Bigg\}.
\eea
Here we have used that the QECC decoding is a linear operation.
The $\EE_{k_2}$ expectation gives rise to a CV twirl, which we evaluate using Lemma \ref{CV-twirl}.
We treat the Gaussian factor in the result of the Lemma as a Dirac delta function.
This yields
\bea
&&\mathcal{C}^{MR \to MRF}(\rho_{MR}) \propto  \Tr_{\rm trap}\EE_{k_1} \int \rd^2\vec{\alpha} |\chi(\vec{\alpha})|^2\Bigg\{ \nn \\
    &&\text{Dec}   \Bigg( \sqrt{V^{\acc}_\epsilon} \pi_{k_1}^\dagger  (D_C(\vec{\alpha}) \otimes U_{R}^{\vec{\alpha}}) \pi_{k_1}\qps\pi_{k_1}^\dagger  (D_C(\vec{\alpha}) \otimes U_{R}^{\vec{\alpha}})^\dagger  \pi_{k_1} 
    \sqrt{V^{\acc\dagger}_\epsilon} \Bigg) \otimes\ket{\text{acc}}\bra{\text{acc}} \nn \\
    &&+\Omega_M \Tr_{M} \Bigg( \sqrt{V^{\rej}_\epsilon} \pi_{k_1}^\dagger  (D_C(\vec{\alpha}) \otimes U_{R}^{\vec{\alpha}}) \pi_{k_1}\qps\pi_{k_1}^\dagger  (D_C(\vec{\alpha}) \otimes U_{R}^{\vec{\alpha}})^\dagger \pi_{k_1} \sqrt{V^{\rej\dagger}_\epsilon} \Bigg)\otimes \ket{\text{rej} } \bra{\text{rej}} \nn \Bigg\}.
\eea
Next we rewrite the permutation of the displacement $D_C(\vec\qa)$ as a displacement over the permuted~$\vec\qa$. 
\bea
&&\mathcal{C}^{MR \to MRF}(\rho_{MR}) \propto  \Tr_{\rm trap}\EE_{k_1}\int \rd^2\vec{\alpha} |\chi(\vec{\alpha})|^2 \Bigg\{ \nn \\
    &&\text{Dec}   \Bigg( \sqrt{V^{\acc}_\epsilon}   (D_C(\pi^{-1}_{k_1}\vec{\alpha}) \otimes U_{R}^{\vec{\alpha}}) \qps (D_C(\pi^{-1}_{k_1}\vec{\alpha}) \otimes U_{R}^{\vec{\alpha}})^\dagger  
    \sqrt{V^{\acc\dagger}_\epsilon} \Bigg) \otimes\ket{\text{acc}}\bra{\text{acc}} \nn \\
    &&+\Omega_M \Tr_{M} \Bigg( \sqrt{V^{\rej}_\epsilon}  (D_C(\pi^{-1}_{k_1}\vec{\alpha}) \otimes U_{R}^{\vec{\alpha}}) \qps  (D_C(\pi^{-1}_{k_1}\vec{\alpha}) \otimes U_{R}^{\vec{\alpha}})^\dagger  \sqrt{V^{\rej\dagger}_\epsilon} \Bigg)\otimes \ket{\text{rej} } \bra{\text{rej}} \nn \Bigg\}.
\eea
Next we explicitly write out the POVM V as specified in (\ref{POVM_V}).
For the C register we use label `msg' for the first $n$ modes, the label `X' for the $z$ trap modes after that,
and `P' for the final $z$ modes.
For conciseness we write only the Accept part.
The Reject part is analogous, and will be presented explicitly again at the end of the analysis.
\bea
&&\mathcal{E}^{MR \to MRF}(\rho_{MR}) \propto  \Tr_{\rm trap}\EE_{k_1} \int \rd^2\vec{\alpha} |\chi(\vec{\alpha})|^2 \Bigg\{ \nn \\
    &&\text{Dec}  \Bigg(  (\one^{\otimes n} \otimes 
    \Bigg[ \int_{-\epsilon}^\epsilon\!\! \rd x \, \ketbra x \Bigg]^{\otimes z} 
    \otimes 
    \Bigg[ \int_{-\epsilon}^\epsilon\!\! \rd p \, \ketbra p \Bigg]^{\otimes z} )
    \nn \\ 
    &&({[D_C(\pi_{k_1}^{-1}\vec{\alpha})]}_{\rm msg} \otimes {[D_C(\pi_{k_1}^{-1}\vec{\alpha})]}_X \otimes {[D_C(\pi_{k_1}^{-1}\vec{\alpha})]}_P \otimes U_{R}^{\vec{\alpha}}) (\text{Enc}_M(\rho_{MR}) \otimes\ketbra X^{\otimes z} \otimes\ketbra P^{\otimes z})  \nn \\
    &&({[D_C(\pi_{k_1}^{-1}\vec{\alpha})]}_{\rm msg} \otimes {[D_C(\pi_{k_1}^{-1}\vec{\alpha})]}_X \otimes {[D_C(\pi_{k_1}^{-1}\vec{\alpha})]}_P \otimes U_{R}^{\vec{\alpha}})^\dagger   
    \Bigg) \otimes\ket{\text{acc}}\bra{\text{acc}} \nn \\
    &&+\mbox{Reject part}
\eea
Next we evaluate the trace over all the trap modes.
In each trap mode independently we get an $x$-integral or $p$-integral
of a displaced squeezed state, with integration interval $(-\epsilon,\epsilon)$, i.e.
an integral of the form $\int_{-\epsilon}^\epsilon \! \rd x \, |\bra{x} D(\qb) \ket X|^2$ for some $\qb\in\CC$. 
It holds that
$|\bra{x} D(\qb) \ket X|^2 = (2\pi e^{-r})^{-1/2}\exp(-\frac{(x-\sqrt2 \Re \qb)^2}{2e^{-r}})$
and
$|\bra{p} D(\qb) \ket P|^2 = (2\pi e^{-r})^{-1/2}\exp(-\frac{(p-\sqrt2 \Im \qb)^2}{2e^{-r}})$.
We get
\bea
    g_1(\qb) &\isdef& \int_{-\epsilon}^\epsilon \! \rd x \, |\bra{x} D(\qb) \ket X|^2
    = \frac12{\rm Erf}\frac{e^{r/2}(\epsilon+\sqrt2 \Re \qb)}{\sqrt2}
    +\frac12{\rm Erf}\frac{e^{r/2}(\epsilon-\sqrt2 \Re \qb)}{\sqrt2}
    \\ 
    g_2(\qb) &\isdef& \int_{-\epsilon}^\epsilon \! \rd p \, |\bra{p} D(\qb) \ket P|^2
    = \frac12{\rm Erf}\frac{e^{r/2}(\epsilon+\sqrt2 \Im \qb)}{\sqrt2}
    +\frac12{\rm Erf}\frac{e^{r/2}(\epsilon-\sqrt2 \Im \qb)}{\sqrt2}
\eea
For $r\gg 1$ and properly tuned $\epsilon$ ($\epsilon > e^{-r/2}$), 
this combination of error functions acts as a selection function that 
equals (almost) 1 if $| {\rm displacement} | \leq \epsilon$  and (almost) 0 otherwise.
The product of all the contributions from the trap states yields an overall selection function $G$,
\be
    G(\pi,\vec\qa) \isdef \prod_{j=1}^z g_1\left( [\pi^{-1}\vec\qa]_{X_j} \right) \;
    g_2\left( [\pi^{-1}\vec\qa]_{P_j} \right).
\ee
Finally, we can write the real-world channel as
\bea
&&\mathcal{C}^{MR \to MRF}(\rho_{MR}) =  \EE_{k_1} \int \rd^2\vec{\alpha} |\chi(\vec{\alpha})|^2
\Bigg\{  G(\pi_{k_1},\vec{\alpha}) \nn \\ 
    &&\text{Dec}  \Bigg(({[D_C(\pi_{k_1}^{-1}\vec{\alpha})]}_{\rm msg} \otimes U_{R}^{\vec{\alpha}})\text{Enc}_M(\rho_{MR})({[D_C(\pi_{k_1}^{-1}\vec{\alpha})]}_{\rm msg} \otimes U_{R}^{\vec{\alpha}})^\dagger\Bigg)
     \otimes\ket{\text{acc}}\bra{\text{acc}} \nn \\
     &&+ [1-G(\pi_{k_1},\vec{\alpha})]   \Omega_M \Tr_{M} {\rm Dec}\Bigg(({[D_C(\pi_{k_1}^{-1}\vec{\alpha})]}_{\rm msg} \otimes U_{R}^{\vec{\alpha}})\text{Enc}_M(\rho_{MR})({[D_C(\pi_{k_1}^{-1}\vec{\alpha})]}_{\rm msg} \otimes U_{R}^{\vec{\alpha}})^\dagger \Bigg) \nn \\
     &&\otimes  \ket{\text{rej} } \bra{\text{rej}} \Bigg\}
\eea

\subsection{The ideal channel}
\label{sec:ideal}

We now specify the ideal channel (\ref{Fgeneral}) for our scheme, again closely following \cite{broadbent2016efficient}.
The register $C$ contains one side of $n+2z$ EPR pairs.
(The other side is denoted as $C'$.)
The attack is applied to the $k_1$-permuted modes;
then the modes are unpermuted and finally it is verified if the EPR pairs are 
unmodified.
Specifically, the simulator checks if more than $t$ modes out of the first $n$ have been noticeably displaced,
and if any of the trap modes have been displaced by more than~$\qd$.

Note that in the CV setting a `standard' EPR pair is given by the two-mode squeezed vacuum
\be
    \ket{\rm EPR} = \frac1{\pi\sinh s}\int\! \rd^2 \qa \; e^{-|\qa|^2(\frac1{\tanh s}-1)}\ket{\qa, \qa^*}
\ee
where $\ket{\qa, \qa^*}$ stands for the tensor product $\ket\qa\otimes\ket{\qa^*}$ of two coherent states.
(see e.g. \cite{jeong2000dynamics}).
The `quality' parameter $s$ determines the amount of entanglement between the two modes.
At $s\to\infty$ there is perfect correlation between their $x$-quadratures and
perfect anticorrelation between their $p$-quadratures.
We will work in the limit $s\to\infty$.
The above state $\ket{\rm EPR}$ can be represented as a 50/50 beamsplitter mixture of two individual squeezed vacuums,
one squeezed in the $x$-direction and one in the $p$-direction.
At $s\to\infty$ these become the $\ket{x=0}$ and $\ket{p=0}$ eigenstate respectively. 
A displaced EPR state $\ket{{\rm EPR}(\qb)}$, with $\qb\in\CC$, is created by mixing
$\ket{x=\Re \qb}$ with $\ket{p=\Im \qb}$.
Since $\{\ket x\}_{x\in\RR}$ and $\{\ket p\}_{p\in\RR}$ are single-mode orthogonal bases,
the states $\{\ket{{\rm EPR}(\qb)}\}_{\qb\in\CC}$ form an orthogonal basis of the two-mode Hilbert space.
This is the equivalent of the four Bell states in DV.
As in the DV case, we can map one `Bell' basis state into another by applying a QOTP encryption
(displacement) to one side of the EPR pair.
Thus it holds that
\be
    2\int_\CC\! \rd^2\qb\,  D_C(\qb) \ketbra\EPR_{CC'} D_C(\qb)\dagg=\one_C\otimes \one_{C'},
\label{EPRbasis}
\ee
where the subscripts $C,C'$ label the two modes.
Next we look at the verification step.
For displacement $u\in\CC^n$
we define a `Hamming weight' $w_\qd(\vec u) = \#\{j|\;\; |u_j|>\qd  \}$ which counts how many of the $n$ modes
have a noticeable displacement.
The set of displacements that get accepted by the simulator is given by
\be
    \cD_\cF\isdef \{ (\vec u,\vec\qg,\vec\qj)\in\CC^{n+z+z}|\;\; 
    w_\qd(\vec u)\leq t  \;\wedge\;  \forall_i |\Re\qg_i|\leq \frac\epsilon{\sqrt2}\;\wedge\;  
    \forall_i |\Im\qj_i|\leq \frac\epsilon{\sqrt2}  \}.
\ee
The simulator's POVM for the verification is written as 
$(V^\acc_\cF, V^\rej_\cF)$, with $V^\rej_\cF=\one-V^\acc_\cF$.
We have 
\be
    V^\acc_\cF = 2^{n+2z}\int_{\cD_\cF}\!\!\!\!\! \rd^2\vec\qb\; 
    D_C(\vec\qb)\ketbra{{\rm EPR}}_{CC'}^{\otimes(n+2z)}D_C\dagg(\vec\qb).
\label{POVMF}
\ee
The mapping that represents the ideal channel is given by

\begin{align}
\label{Fmap1}
    &\mathcal{F}^{MR \to MRF}: \rho_{MR} \mapsto 
    \Tr_{CC'} \EE_{\pi \in \cS_{n+2z}} \Big\{\\
    &\Big( \sqrt{V^\acc_{\cF}} \pi_C^\dagger U_{CR} \pi_C 
    (\rho_{MR} \otimes \ketbra{{\rm EPR}}^{\otimes(n+2z)}_{CC'})
    \pi_C^{\dagger}  U_{CR}^{\dagger} \pi_C\sqrt{{V^{\acc\dagger}_{\cF}}} \Big) 
    \otimes\ket{\text{acc}}\bra{\text{acc}} \nn\\
    &+  \Omega_M \Tr_{M}  \Big(  
   \sqrt{V^\rej_{\cF}}  \pi_C\dagg U_{CR} \pi_C 
    (\rho_{MR} \otimes \ketbra{{\rm EPR}}^{\otimes(n+2z)}_{CC'})
    \pi_C^{\dagger}  U_{CR}^{\dagger} \pi_C \sqrt{{V^{\rej\dagger}_{\cF}}} \Big)
    \otimes \ket{\text{rej} } \bra{\text{rej}}
    \Big\}.
    \nn
\end{align}
Here $\cS_{n+2z}$ stands for the set of permutations of $n+2z$ modes.
Again we write
$U_{CR} = \int \rd^2\alpha_1...\rd^2\alpha_{n+2z} \, \chi(\vec{\alpha}) \, D_C(\vec{\alpha}) \otimes U_{R}^{\vec{\alpha}}$ with normalisation 
$\int \rd^2\alpha_1...\rd^2\alpha_{n+2z} \, |\chi(\vec{\alpha})|^2 = 1$.
Again we use $\pi\dagg D(\vec\qa)\pi=D(\pi^{-1}\vec\qa)$.
Furthermore we rotate $\sqrt{V_\cF}$ under the $CC'$-trace so that the square roots combine into $V_\cF$;
then we substitute the POVM (\ref{POVMF}) into (\ref{Fmap1}).
This gives
\bea
    & \mathcal{F}^{MR \to MRF}(\qr_{MR}) \propto 
    \EE_{\pi\in\cS_{n+2z}} \int\rd^2\vec\qa \rd^2\vec\qa' \chi(\vec\qa) \overline{\chi(\vec\qa')} \Tr_{CC'}
    \int \rd^2\vec\qb \Big\{ \nn\\ &
    I(\qb\in\cD_\cF) U^\qa_R \qr_{MR} U_R^{\qa'\dagger} \otimes D_{\pi^{-1}\vec\qa'}\dagg D_\qb \ketbra{{\rm EPR}}^{\otimes(n+2z)}_{CC'}
    D_\qb\dagg D_{\pi^{-1}\vec\qa} \ketbra{{\rm EPR}}^{\otimes(n+2z)}_{CC'}
    \otimes \ketbra{\acc}
    \nn\\ &
    + I(\qb\notin\cD_\cF) \qO_M\otimes {\rm Tr}_M [\cdots{\rm same}\cdots] \otimes \ketbra\rej
    \Big\}
\eea
Here all the displacements act on the $C$ space;
the $I(\qb\in\cD_\cF)$ is an indicator function that equals 1 when the condition is met;
the abbreviation `same' stands for the same state in $MRCC'$ space as in the line above.
Since the EPR states count as beamsplitter-mixtures of perfect $x$- or $p$-eigenstates,
the trace ${\rm Tr}_{CC'}$ acting on the displaced EPR states yields a product of Dirac delta functions,
$\qd(\vec\qb-\pi^{-1}\vec\qa) \qd(\vec\qb-\pi^{-1}\vec\qa')$,
which can be rewritten as
$\qd(\vec\qa'-\vec\qa) \qd(\vec\qb-\pi^{-1}\vec\qa)$.
Carrying out the integrals over $\vec\qa'$ and $\vec\qb$ yields
\bea
    \mathcal{F}^{MR \to MRF}(\qr_{MR}) &=&
    \EE_{\pi\in\cS_{n+2z}} \int\rd^2\vec\qa |\chi(\vec\qa)|^2 \Big\{
    I(\pi^{-1}\vec\qa\in\cD_\cF)U_R^\qa\qr_{MR}U_R^{\qa\dagger}\otimes\ketbra\acc
    \nn\\ && +
    I(\pi^{-1}\vec\qa\notin\cD_\cF) \qO_M \otimes {\rm Tr}_M U_R^\qa\qr_{MR}U_R^{\qa\dagger}
    \otimes\ketbra\rej\Big\}.
\eea

\subsection{Finishing the proof}

Note that we can write 
$I(\pi^{-1}\vec\qa\in\cD_\cF)U_R^\qa\qr_{MR}U_R^{\qa\dagger}$
in the more complicated form
$I(\pi^{-1}\vec\qa\in\cD_\cF) {\rm Dec}\Big( [D_{\pi^{-1}\vec\qa}]_{\rm msg}\otimes U_R^\qa 
\;\; {\rm Enc}\qr_{MR}\;\; [D_{\pi^{-1}\vec\qa}]_{\rm msg}\dagg\otimes U_R^{\qa\dagger}\Big)$.
This equality holds because, under the condition on $\vec\qa$, the decoding is guaranteed to recover $\qr_{MR}$.
We use the more complicated form to express the difference $\cC-\cF$ in a compact form,
\bea
    & \cC(\qr_{MR}) - \cF(\qr_{MR}) =
    \EE_{\pi\in\cS_{n+2z}} \int\rd^2\vec\qa \; |\chi(\vec\qa)|^2 \Big[ \Big\{ G(\pi,\vec\qa)-I(\pi^{-1}\vec\qa\in\cD_\cF) \Big\}
    \nn\\ &
    {\rm Dec}\Big( [D_{\pi^{-1}\vec\qa}]_{\rm msg}\otimes U_R^\qa 
    \;\; {\rm Enc}\qr_{MR}\;\; [D_{\pi^{-1}\vec\qa}]_{\rm msg}\dagg\otimes U_R^{\qa\dagger}\Big)
    \otimes\ketbra\acc
    \nn\\ &
    + \Big\{ 1-G(\pi,\vec\qa)-I(\pi^{-1}\vec\qa\notin\cD_\cF) \Big\}\qO_M \otimes{\rm Tr}_M {\rm Dec}(\cdots{\rm same}\cdots)
    \otimes\ketbra\rej \Big]
    \\ & =
    \EE_{\pi\in\cS_{n+2z}} \int\rd^2\vec\qa \; |\chi(\vec\qa)|^2 \Big\{ G(\pi,\vec\qa)-I(\pi^{-1}\vec\qa\in\cD_\cF) \Big\}
    \nn\\ & \Big\{
    {\rm Dec}\Big( [D_{\pi^{-1}\vec\qa}]_{\rm msg}\otimes U_R^\qa 
    \;\; {\rm Enc}\qr_{MR}\;\; [D_{\pi^{-1}\vec\qa}]_{\rm msg}\dagg\otimes U_R^{\qa\dagger}\Big)
    \otimes\ketbra\acc
    \nn\\ &
    - \qO_M \otimes{\rm Tr}_M {\rm Dec}(\cdots{\rm same}\cdots)
    \otimes\ketbra\rej \Big\}.
\eea
Next we use the triangle inequality to obtain the following bound
\bea
    & \| \cC(\qr_{MR}) - \cF(\qr_{MR}) \|_{\rm tr} \leq \EE_{\pi\in\cS_{n+2z}} \int\rd^2\vec\qa \; |\chi(\vec\qa)|^2 \Big\{ G(\pi,\vec\qa)-I(\pi^{-1}\vec\qa\in\cD_\cF) \Big\}
    \nn\\ &
    \Big\| 
    {\rm Dec}\Big( [D_{\pi^{-1}\vec\qa}]_{\rm msg}\otimes U_R^\qa 
    \;\; {\rm Enc}\qr_{MR}\;\; [D_{\pi^{-1}\vec\qa}]_{\rm msg}\dagg\otimes U_R^{\qa\dagger}\Big)
    \otimes\ketbra\acc
    \nn\\ &
    - \qO_M \otimes{\rm Tr}_M {\rm Dec}(\cdots{\rm same}\cdots)
    \otimes\ketbra\rej  \Big\|_{\rm tr}
    \\ &
    \leq  \int\rd^2\vec\qa \; |\chi(\vec\qa)|^2 \EE_{\pi\in\cS_{n+2z}} \Big\{ G(\pi,\vec\qa)-I(\pi^{-1}\vec\qa\in\cD_\cF) \Big\}.
\label{GIexpression1}
\eea
In the last step we used that the trace distance between two normalised states cannot exceed~1.

Note that the functions $G$ and $I$ are very similar.
The indicator $I$ exactly selects displacements $\vec\qa\in\CC^{n+2z}$
such that in the traps part of $\pi^{-1}\vec\qa$ the measured component is $\frac\epsilon{\sqrt2}$-close to zero,
and in the message part of $\pi^{-1}\vec\qa$ the Hamming weight $w_\qd$ is low.

The function $G$ is not an exact indicator function, having continuous behaviour.
However, for $e^{-r/2}\ll \epsilon$ it is extremely close to a step function;
we will assume that we are in this regime.
The $G$ enforces the same conditions as $I$ on the traps part of $\pi^{-1}\vec\qa$,
but ignores the message part.

The expression $G(\pi,\vec\qa)-I(\pi^{-1}\vec\qa\in\cD_\cF)$ evaluates either to 0 or 1;
it cannot become negative since $I$ imposes more conditions than~$G$.
The value 1 occurs only if the traps are intact but the message has uncorrectable noise.
(See Table~\ref{tab:GminusI}).

\begin{table}[h]
\centering
\begin{tabular}{|l|c|c|c|}
\hline
\textbf{Case} & \( G \) & \( I \) & \( G - I \) \\
\hline
All modes have negligible displacement & 1 & 1 & 0 \\
Some trap has too much displacement & 0 & 0 & 0 \\
All traps OK, message not OK (uncorrectable error) & 1 & 0 & 1 \\
\hline
\end{tabular}
\caption{Behavior of indicator functions \( G \), \( I \), and their difference in different attack scenarios.}
\label{tab:GminusI}
\end{table}

For the final step in the proof we have to tune the parameter $\qd$ to $\qd=\frac\epsilon{\sqrt2}$
in order to obtain symmetry between all the modes.
Let $u$ be the number of modes in $\vec\qa$ that contain a large displacement.
We consider only vectors $\vec\qa$ that can yield $G-I=1$ for some permutation~$\pi$.
Such a vector must have $u\in\{t+1,\ldots,n\}$.
The expression $\EE_{\pi\in\cS_{n+2z}} (G-I)$ is the probability, 
given a random permutation of $n+2z$ modes, of placing the $u$ noisy ones precisely
in the first $n$ positions. This probability is given by
\be
    P(u) = \frac{{n \choose u} u!}{{n+2z \choose u} u!} = \frac{n!}{(n+2z)!}(n-u+1)\cdots(n-u+2z).
\label{combiprob}
\ee
As $P(u)$ is a decreasing function of $u$ we can write $P(u) \leq P(t+1)$.
Next we write
\be
    P(t+1)=\frac{{n\choose t+1}}{{n+2z\choose t+1}}= \prod_{j=0}^t\frac{n-j}{n+2z-j}
    <  \prod_{j=0}^t \frac{n}{n+2z} = (\frac{n}{n+2z})^{t+1}.
\ee
Here we have used the inequality 
$\frac{n-j}{n+2z-j} \leq   \frac{n}{n+2z}$, which holds for $2z> n$.

Finally we use the normalisation of $\chi$ and obtain the end result
\be
    \forall_{\qr_{MR}}\quad  \| \cC(\qr_{MR}) - \cF(\qr_{MR}) \|_{\rm tr} < (\frac{n}{n+2z})^{t+1}.
\ee
Hence we satisfy the security definition (\ref{def:security}) with $\qh=(\frac{n}{n+2z})^{t+1}$.
This is very similar to the DV result $(\frac13)^{t+1}$ in~\cite{broadbent2016efficient},
but with flexibility in the number of traps.


\section{Discussion}
\label{sec:discussion}

Our continuous-variable construction and its security proof bring no real surprises to those familiar with
the discrete-variable quantum authentication schemes.
However, some technical hurdles had to be overcome, e.g.\;dealing with the non-perfect CV QOTP, introducing the CV twirl 
and handling the approximate step functions.
A more rigorous treatment of the approximate step functions (in which the difference between the $I$ and $G$ indicators
ends up as a small addition to $\qh$) is left for future work.

Note that the scheme authenticates a single-mode state. 
This is readily generalized to multiple modes either by authenticating each mode individually or
by applying a quantum error-correcting code to a multi-mode message.
We note that
our counting argument (\ref{combiprob}) is presented in a bit more direct way 
than the derivation in the Appendix of~\cite{broadbent2016efficient}.

\bmhead{Acknowledgments}
Part of this work was supported by the Dutch Startimpuls NAQT CAT-2 and NGF Quantum Delta NL CAT-2.




\bibliography{CV_Auth}

\end{document}